\documentclass[twocolumn]{article}

\usepackage[english]{babel}
\usepackage{microtype}
\usepackage{float}
\usepackage{biblatex}
\usepackage[letterpaper,top=2cm,bottom=2cm,left=3cm,right=3cm,marginparwidth=1.75cm]{geometry}

\usepackage{amsmath}
\usepackage{graphicx}
\usepackage[colorlinks=true, allcolors=blue]{hyperref}
\usepackage{booktabs}

\title{Simplifying Cyber Cat(astrophe)s with Cyber Kittens: \newline
Power Law Plausibility for Cyber Insurance Risks}
\author{Max Henderson, Anton Solomko, Henry Simmons, \\Nick Jin, Maximilian Kloucek, and John Wingate}

\begin{document}

\maketitle
\begin{abstract}
Cyber insurance requires accurate modeling of worst-case catastrophic (cat) events, but the field lacks robust quantitative approaches for estimating upper-bound losses. Building on a recent dataset of 24 cyber cat events over 30 years \cite{johansmeyer2026rewriting}, this work tests whether cyber economic losses follow a power law distribution. We analyze ``cyber kittens"—sub-\$1B events distinguished from cat events (\$1B+) only by magnitude—extracted via LLM from cyber insurance claims data (2020–2024). Using victim count (weighted by claim year) as a proxy for economic loss, we link kitten-sized events to known cat events to estimate losses. The kitten distribution proved consistent with the cat dataset, and power laws were statistically plausible: each order-of-magnitude increase in event size corresponds to a 5–7x drop in probability. Extrapolating, an event 100x the largest 2020–2024 cat event is expected roughly every 206 years, translating to \$100B–\$250B in losses—catastrophic, but not extraordinary relative to other insurance lines.
\end{abstract}

\section{Introduction}\label{introduction}

A common observation is that the world is becoming more technological, and that, if anything, the pace of this digitization is increasing. A concrete example of this were the digital shifts experienced during the Covid pandemic, ranging from student education moving online \cite{Koh2022, Ives2021} to workplaces opting for work-from-home solutions such as Zoom \cite{Barrero2025, Center2022}.

Unsurprisingly, these shifts in technological usage are not without tradeoffs. In the Covid example, as many soft targets joined or expanded their digital attack surface, hackers leapt at the opportunity. This was clearly observed by the cyber insurance community, who provide financial risk transfer products to businesses, government bodies, and other organizations as well as private individuals. These products have a wide range of coverage areas, including extortion, business interruption, and funds transfer fraud. Covid was seen as an accelerant of the ransomware trend that began in 2020, and this trend resulted in the highest losses the cyber insurance world had yet seen \cite{CyberMaxx2024, Ratings2025, Coalition2020, Tsohou2023, Holdings2025}, causing insurers to increase security requirements, declinations, and premium prices.

The cyber insurance ecosystem rebounded, but the losses experienced during this period were comparatively mild to the existential concerns of the market. While 2020--2022 presented a few challenging years from an insurance perspective, these were ultimately nothing more than a set of higher-than-average single company losses, due to higher frequency and severity of ransomware events. The true concerns of the market revolve around cyber catastrophes, or cyber cat events. Cyber cat events are cyber incidents that generate a large number of correlated losses. An example of such an event is a coordinated distributed denial of service (DDoS) attack that leads to a large portion of the entire global network of companies suffering business interruption. The exact definition of a cyber cat event varies across different organizations and academics, but typically defines a cyber cat as an event that surpasses either an insured or economic loss threshold. Organizations like PCS and PERILS define cyber cats using insured loss, with thresholds of \$250M (globally) and \$500M (United States only), respectively \cite{pcs2018globalcyber, perils2026cyber}.The goal of \cite{johansmeyer2026rewriting} was to publish a comprehensive set of cyber cat events over the last 30 years with their corresponding best estimates of economic losses, which were formulated using a clear and robust evaluation methodology. The work's definition of a cyber cat event used was an economic loss threshold of \$800M in a pragmatic effort to maximize the size of an dataset, given the extremely small number of such events.

 Quantifying the frequency and severity of these kind of systemic cyber risks is the primary work of the cyber reinsurance community, and this risk estimation forms the basis for capacity requirements that support all cyber insurance products. A challenge for cyber is that there remains a strong and underlying fear that cyber risk presents a truly novel kind of peril to the market. Cyber insurance has a notably high rate of cession to reinsurers compared with other lines of business, reflecting carriers' discomfort with retaining cyber risk. Modeling cyber cat events does not have a consensus view, as shown in a report from Axis Insurance \cite{Donald2019}. Is systemic risk best modeled as a kind of immune system with feedback loops, physical models with catataxic shifts, or complex causal models? Many other perils do seem to break qualitatively from cyber, due to its inherent complex global interconnection. A hurricane in southern Florida will have minimal impact on California, but can the same be confidently said for cyber risk vectors?

This highlights another challenge for quantifying cyber cat events: the long-standing dearth of data across the industry. Carriers have sufficient data to price attritional risks using firmographic details such as industry and company size. However, a study by \cite{Cremer2022} showed that over 5,000 peer-reviewed papers on cyber risk used fewer than 100 unique datasets. A small number of trusted cyber datasets makes complex causal modeling approaches for as-of yet unseen cyber cat events essentially intractable. While the recent dataset containing 24 cat events contained in \cite{johansmeyer2026rewriting} provides a meaningful contribution to the space, additional analysis is clearly need, as the work notes that ``...it  provides a desperately needed starting point for further study of catastrophic cyber events."

Our work builds on this quantitative research direction by evaluating a simple hypothesis: that the distribution of cyber cat events can be plausibly explained using power laws. Power law relationships, wherein there is a linear relationship between log(frequency) and log(magnitude), have been observed across a wide range of phenomena, including earthquakes \cite{Gutenberg1944}, wars and conflicts \cite{Clauset2007, Richardson1948}, populations of cities \cite{Gabaix1999}, paper citations \cite{Redner1998}, and power outages \cite{Balint2023}. These simple functional relationships are powerful and can help better parametrize ``gray swan" events \cite{Taleb2007}, e.g., having a strong confidence in the frequency of observing an earthquake of magnitude 8 in the Philadelphian region, despite never having observed one in the past. Critically, power laws can make these predictions on macroscopic properties of systems, without the need for additional information or context. 

\begin{figure*}[h]
    \centering
    \includegraphics[width=0.99\linewidth]{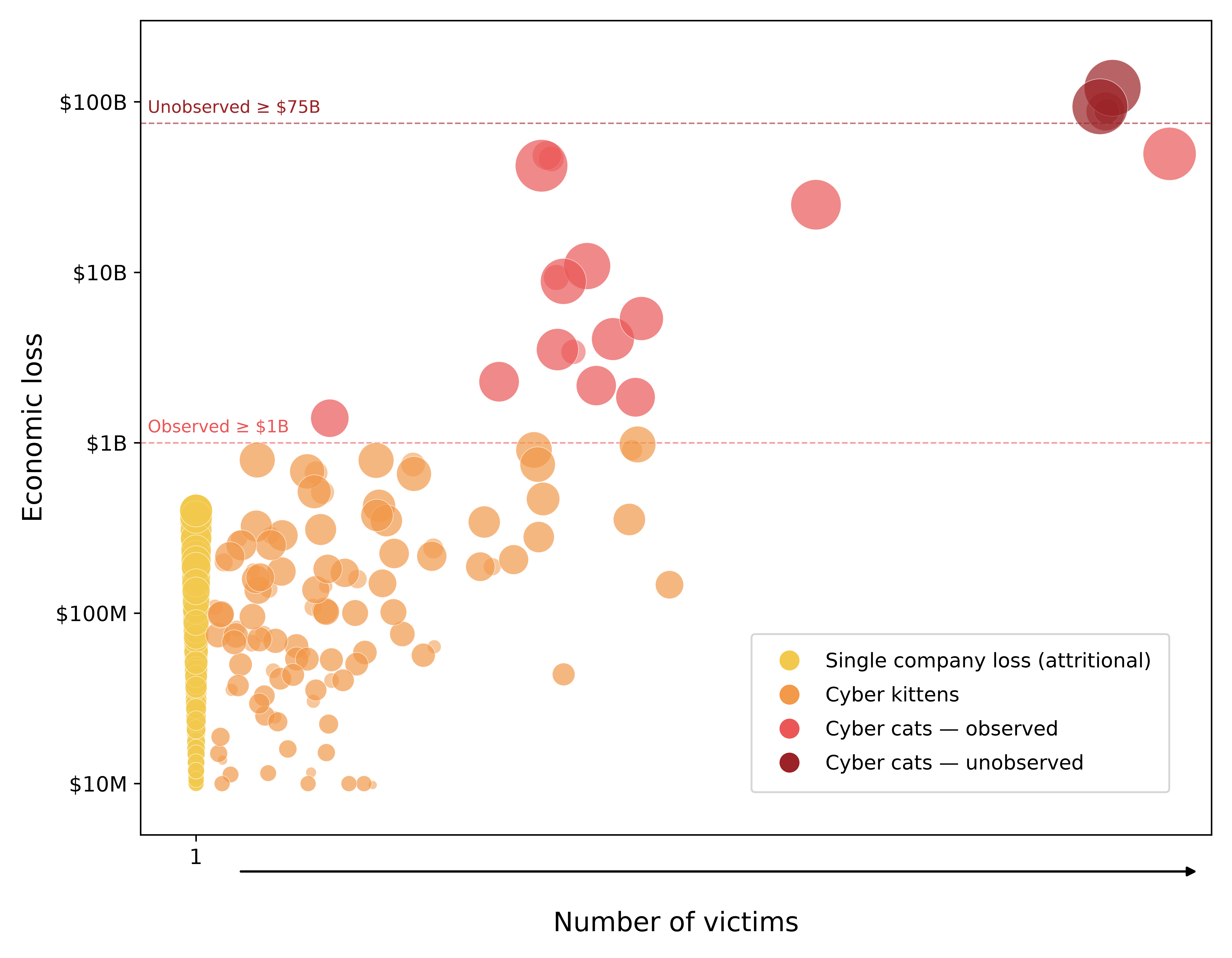}
    \caption{Visualizing a synthetic data distribution of single company losses, and the transition from cyber kitten to cyber cat events.}
    \label{kittens_and_cats}
\end{figure*}

Power laws are also not new to the space of cyber risk. Power law behavior has been observed for both frequency and severity, with empirical evidence from data breaches and ransomware showing heavy-tailed distributions in which extreme events dominate aggregate losses \cite{Maillart2010, Edwards2016, Leverett2020}. However, empirical studies that better quantify the estimates of power law behavior for large cyber cats have not been published anywhere in the academic literature to date. Fundamentally, performing such analysis requires sufficient data; to properly test if economic losses due to cyber cats are power law distributed, one would need many more samples than available in the literature. Although \cite{johansmeyer2026rewriting} deliberately choose an economic threshold to maximize the number of events to consider, this results in a dataset of 6 events over the last decade, which is infeasible for assessing power law (or other distributional) fits.

The primary contribution of this research is enable proper power law analysis on cyber cat events by forming a bridge with cyber kitten events, which can be visualized in Fig. \ref{kittens_and_cats}. We believe cyber cats and cyber kittens follow the same generating distribution, and define both as cyber events in which there are multiple unique victim organizations, which we would measure explicitly using cyber insurance claims. The primary difference between the two uses an economic threshold similar to \cite{johansmeyer2026rewriting}: just as kittens are smaller than cats, we define cyber kittens as economic losses below \$1B, while cyber cats are \$1B or higher.

This plot visualizes several aspects of the assumptions underlying this work. First, is that while single company losses can be larger than cyber kittens, the largest losses we have seen (and could see) would necessarily be connected to events involving more than 1 victim. The Marriott Hotel hack and Capital One data breach are two of the largest known single company cyber losses, likely exhausting insurance towers of \$350M and \$400M, respectively \cite{artemis2018marriott, reinsurancenews2018marriott, artemis2019capitalone, reinsurancenews2019capitalone}. The economic losses of large single company losses can be considerably larger than many cyber kittens, but these kind of events have been well within the insurance market's ability to absorb. The true ``tail risk" comes from a large number of companies experiencing losses that drive the summed economic loss into previously unobserved territories. Our second main assumption, is the number of victims is correlated with the overall economic loss of each event. We assume this is a noisy functional relationship, wherein some events with identical victim counts will have economic losses than can vary by an order of magnitude. But we presume that this still leads to a transition wherein most events above a certain victim count threshold end up being cyber cats rather than kittens. Finally, the key move in this work is in asserting that the cyber kittens and cats follow from the same underlying distribution, which we hypothesize to be a power law. By showing plausibility of a power law to describe the entire distribution of cyber cats and kittens, as well as empirical evidence for a transition from cyber kittens to cats, we can speculate on cyber cat event size frequency beyond what has been previously observed.

Using a set of cyber insurance claims from 2020--2024, we extract a dataset of over 150 cyber kitten events, as well as the 4 cyber cat events that occurred during this period (from table in \cite{johansmeyer2026rewriting}). This analysis was made possible by leveraging state-of-the-art LLM technologies, which were critical for finding and extracting cyber kittens from our claims dataset. This dataset forms the backbone of the analysis in this work, which is structured as follows. In the Methodology section, we highlight details of our claims dataset, our LLM-enrichment approach that generates our cyber kittens dataset, and our statistical framework for evaluating power law plausibility. Our findings are shown in the Results section, where we show empirical evidence for the plausibility of a power-law relationship for the cyber kittens and the corresponding implications for observing cyber cat events of \$100B losses and beyond. Finally, in the Discussion we speculate on how results like this could help stabilize the cyber insurance market, particularly at a time where discussions on the risk of cyber war loom large.

\section{Methodology}\label{methodology}

Research has moved in a more quantitative direction for measuring cyber risk and extreme events, despite the data limitations for straightforward analysis. An interesting approach for modeling attritional cyber risk without cyber claims or policy data was implemented in \cite{woods2021countyfair}, who essentially reverse-engineered an insurance premium model from regulatory filings of 26 insurers via a particle swarm optimization algorithm. In terms of quantifying worst-case cyber events, a variety of realistic disaster scenarios have been explored, from Sybil logic bombs to attacks on critical energy infrastructure \cite{ruffle2014sybil, kelly2016integrated}. As pointed out by \cite{eling2023economic}, while interesting these approaches lacked common methodologies which perhaps unsurprisingly led to highly variant results, spanning 0.2\% to 2\% of the gross domestic product (GDP). \cite{eling2023economic} then applied a dynamic in-operability input-output model to 6 extreme cyber scenarios and found losses between \$0.7B--\$35B, which while still wide, was squarely within an insurable range. And as already described, outside of quantitative models for cyber risk, the work of \cite{johansmeyer2026rewriting} provided a strong dataset of cyber cat events to the community via a transparent methodological process.

This work continues in this quantitative zeitgeist of analyzing cyber risk, combining a simple mathematical power law model to a relevant cyber insurance claims dataset, of which cyber kitten and cat events have been extracted. At the modeling level, fitting a power law to explain cyber cat dynamics is mechanically no different than many other approaches, whether the fit be done with regression models, neural networks, or some other optimization framework. Philosophically however, the simplicity of the power law and the fundamental lack of additional context does play a bit of devil's advocate to other scenario models, which build out details according to the specific characteristics of each type of loss. A succinct example in a science communication video on power laws by \textit{Veritasium}\footnote{Derek Muller and Casper Mebius, "You've (Likely) Been Playing The Game of Life Wrong," *Veritasium*, YouTube, November 26, 2025, https://www.youtube.com/watch?v=HBluLfX2F\_k} illustrates this concept: despite the immense complexity of the earth -- biological networks, weather patterns, human-made systems -- a simple gravitational equation (i.e., a power law) is all that is required to accurately predict a macroscopic property of the system (e.g., the orbital motion of the earth). The power law methodology makes no claims about in the specifics of future cyber cat events, such as likely hit geographical regions, which industry vertical are most risky, or whether/which nation states would be most likely to generate them. Power law plausibility in this work simply allows us to model one macroscopic property of cyber cat events: the frequency of which we'd expect to observe events of different economic loss magnitudes.

In this section, we review the methodology for testing our power law hypothesis for our cyber kittens dataset. First, we describe the cyber insurance claims data and LLM enrichment for extracting the primary dataset to conduct our analysis. Second, we explore results around our cyber kittens dataset output, defining precisely how we ``weigh" cyber kittens and show this is consistent with known economic losses connected to cyber cats. Third, we review the basic math and statistical approach for testing the plausibility of our power law hypothesis on this extracted dataset.

\subsection{Claims data and LLM enrichment}\label{claims_data}

The analysis in this work is built on a subset of cyber insurance claims from 2020--2024. Critically, the claims in this dataset had several text fields which could be combined to form a meaningful description of the event. An initial sanity check of these descriptions using regex techniques was useful for finding significant claim counts connected to 4 cyber cat events during this period from \cite{johansmeyer2026rewriting}: MOVEit \cite{Emsisoft2023}, Change Healthcare \cite{changehealthcare}, CDK Global \cite{Group2024}, and CrowdStrike \cite{parametrix2024}.

Improving on the regex model, we built a pipeline which leveraged Google Gemini (gemini-3-flash-preview) to label each claim as attritional (i.e., one-off, random, no correlation with other losses) or systemic (i.e., correlated with other losses/claims). If the claim was systemic, we also prompted the LLM to label the loss, and then ran a second order pass on all labels to prevent collisions, as well as manual review with a few hard-coded corrections based on string similarity (i.e., ``CrowdStrike" vs ``CrowdStrike update"). Finally, as we defined cyber kittens as involving more than one victim, we dropped any events that did not have at least 2 unique victims. At the end of this process, we extracted over 150 cyber kitten events connected to over 5,000 unique claims.

To validate the fidelity of this extraction process, we performed some manual validation on a subset of the underlying labeled claims. We randomly selected a comparable set of 300 attritional claims (i.e., LLM labeled as non-systemic) and 300 claims connected to cyber kittens. We saw 4 mislabeled events in the set of ~300 for an error rate of approximately 1\%. Each of the mislabeled data points was an interesting case study for the strengths and limitations of LLM-based approaches on claims. Two claims were systemic but mislabeled, and both originally for the Change Healthcare event. One was actually for the Ascension Healthcare event (likely driven by similar healthcare space descriptions) and the other was a ALPHV / BlackCat ransomware event. Blackcat was attributed to the Change Healthcare event, but the loss was in 2023 (a separate event) while Change Healthcare occurred in 2024. The other two were attritional events with similar text descriptions to other systemic losses, but no identifiable information that would indicate they were part of a specific event. None of the 300 randomly selected attritional claims were found to be mislabeled as systemic. Our manual inspection yielded promising estimates for the precision and recall of LLM-enrichment of cyber insurance claims for systemic event identification.

Finally, to ensure data confidentiality, all queries submitted to the LLM were processed in accordance with the provider's data usage policies, which explicitly state that user-submitted data is not used to retrain or improve their models \cite{Google2026}. This safeguard ensures that sensitive information shared during the enrichment process remains protected and will not be leveraged for third-party model development.

\subsection{Weighing cyber kittens}\label{weighing_cyber_kittens}

As defined in Section \ref{introduction}, the primary difference between cyber cats and kittens is their economic loss, or how much they ``weigh". The challenge for weighing our cyber kittens directly is twofold. First, our claims data contains information on \textit{insured} losses, not direct economic losses. Insured losses are essentially economic losses put through a non-linear ``clipping" function, that obfuscate exact economic losses above the limit and below the deductible. Second, to create the most robust dataset of claims as possible, we analyzed both open and closed claims, the difference being that closed claims have fully resolved the insured loss, while open claims have estimates or placeholder values. 

\begin{figure}[h]
    \centering
    \includegraphics[width=0.99\linewidth]{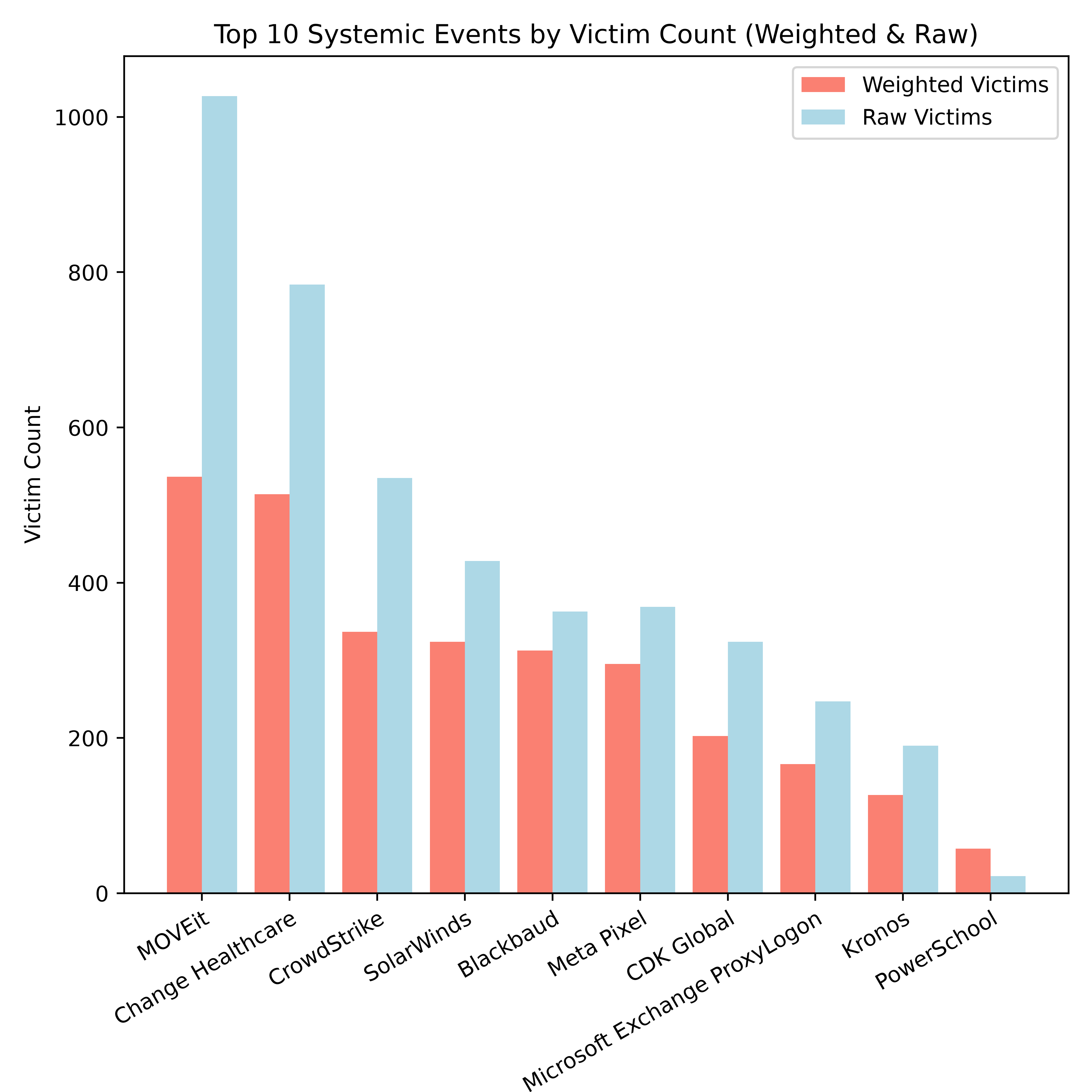}
    \caption{Visualizing cyber kitten raw and weighted victim counts for the top 10 cyber kittens in 2020--2024 cyber claims dataset.}
    \label{top10}
\end{figure}

To work around these issues, the cyber kittens in our dataset were ranked using a weighted victim count, which can be visualized comparatively against raw victim counts in Fig. \ref{top10}. Raw victim counts are simply the count of unique organizations who filed at least one cyber insurance claim connected to a cyber kitten event. The weighting comes from asymmetries in annual number of total claims analyzed; for years with fewer claims, each claim ``counts more" with a higher weight, and years with more claims ``count less" comparatively.

The top 10 events by weighted victim counts in Fig. \ref{top10} validates our general hypothesis proposed in Section \ref{introduction}, and visualized in Fig. \ref{kittens_and_cats}. In a set of 150 cyber kittens, all 4 of the cyber cats enumerated in Section \ref{claims_data} are in the group of top 10 weighted victim counts. As suggested, the relationship between victim count and economic loss is not likely a 1-to-1 mapping; while the top 3 events were all cyber cats, CDK Global was actually ranked as the 7th largest. Nevertheless, the relative ranking shows a clear transition when aggregating by weighted victim count; 60\% of the top 1--5 events were cyber cats, 20\% of the top 6--10 events were cyber cats, and none of the bottom 90\% of other events contained a cyber cat.

Based on our manual vetting of the cyber kitten labels from the LLM procedure coupled with consistent results that link weighted victim counts to economic losses, we can lay out our evaluation framework for the statistical plausibility of a power law to describe the distribution of these weighted victim counts.

\subsection{Power law hypothesis}\label{power_law_hypothesis}

\subsubsection{Definition and motivation for cyber}\label{power_law_def}
The power law equation is simple:
\begin{equation}
\label{pure_power_law}
p(x) \propto x^{-\alpha}
\end{equation}
wherein $p(x)$ is the probability of observing an event of size $x$ and $\alpha$ is a positive constant value greater than 1. Power laws are patterns observed in self organizing critical systems, which produce results with extremely heavy tails (i.e., non-negligible probabilities of observing extremely large outliers). Depending on your system, $x$ could represent the population of a group of cities, earthquake magnitudes, or paper citations. Importantly, power law distributions emerge from graphical networks. This is an intuitive ansantz for cyber risk, which spreads via a global network of companies, governments, and organizations who are connected via shared technologies and supply chains. The Crowdstrike incident is an example of a cyber cat connected such technology connections, as a large number of victims were impacted due to specific Crowdstrike product update. On the supply chain side, WannaCry caused large losses connected to energy and food related supply chain interruptions.

\subsubsection{Statistical methodology for power law fit testing}\label{power_law_testing}

As pointed out by \cite{Clauset2009}, many datasets with heavy tails have been fit empirically to power laws, using simple least-squares fitting techniques. This can lead to many problematic outcomes. High fluctuations in the tail events of such distributions combined with an often incorrect assumption that all data points are actually part of the range in which the power law applies can lead to highly erroneousness best fit power law parameters. Additionally, there are many heavy tailed distributions which may actually describe the underlying data distribution better than a power law. A power law fit to a data distribution might lead to a low least-squares error, but still be inferior to other distributions.

To properly analyze our own cyber-relevant datasets, we used statistical evaluation techniques from \cite{Clauset2009}  that were motivated by a desire to rigorously detect, characterize, and parametrize power law behavior over a variety of empirical datasets. The methods from the paper were implemented by the \texttt{powerlaw} package \cite{Alstott2014}, which we used to do our analysis. To briefly review, the approach for analyzing the fidelity of a power law fit combines various maximum-likelihood fitting methods with goodness-of-fit tests established from the Kolmogorov-Smirnov (KS) statistic and likelihood ratios, detailed in Section 4 of \cite{Clauset2009}. This method fits the data to a power law and a corresponding KS distance. Then, for $N$ simulation rounds, that best fit power law is used to generate synthetic data, which is re-fit to a new power law and respective KS distance. The fraction of the time the resulting statistic is larger than the original KS value for the empirical data is the $p$-value, wherein values above 0.1 support plausibility of the data being power law distributed, and vice versa for values $\leq 0.1$.

\cite{Clauset2009} also provides sound methods of comparisons between a power law and different distributions fit on input data. The primary distribution comparison is done by calculating the normalized log likelihood ratio $R$ and its corresponding p-value, which is the statistical significance of that ratio. If the $R$ values is negative, that means the comparison distribution is a better fit compared to a pure power law, and vice versa if positive, while a $p$-value $\leq$ 0.05 is typically considered to be statistically significant. In this paper we test power laws against power law with cutoff and the lognormal distribution. The power law with cutoff is quite similar to the pure power law equation of Eq.~\ref{pure_power_law}:
\begin{equation} \label{power_law_with_cutoff} 
p(x) \propto x^{-\alpha}e^{-\lambda x}
\end{equation}
where the exponential term is has minimal impact for smaller $x$ events, but as $x$ grows larger, drags the probability of observation toward 0. Mathematically, arbitrarily large values still have positive probabilities of measurement, but in practical scenarios with a finite number of samples, observing values beyond a certain size becomes virtually impossible. Finally, the lognormal distribution equation is as follows:
\begin{equation} \label{lognnormal} 
p(x) \propto \frac{1}{x} e^{-\frac{(\ln x - \mu)^2}{2\sigma^2}},
\end{equation}
wherein $\mu$ and $\sigma$ are the mean and standard deviation of $\ln(x)$, respectively. In practical terms, these three options capture different speeds at which the extreme events decay as a function of $x$. Lognormal distributions decay the fastest, followed by power law with cutoff, while pure power laws remain (by definition) unaffected by the scale of $x$.

\section{Results}\label{results}
\subsection{Power law plausibility for cyber kittens}\label{power_law_results}

\begin{figure}[h]
    \centering
    \includegraphics[width=0.9\linewidth]{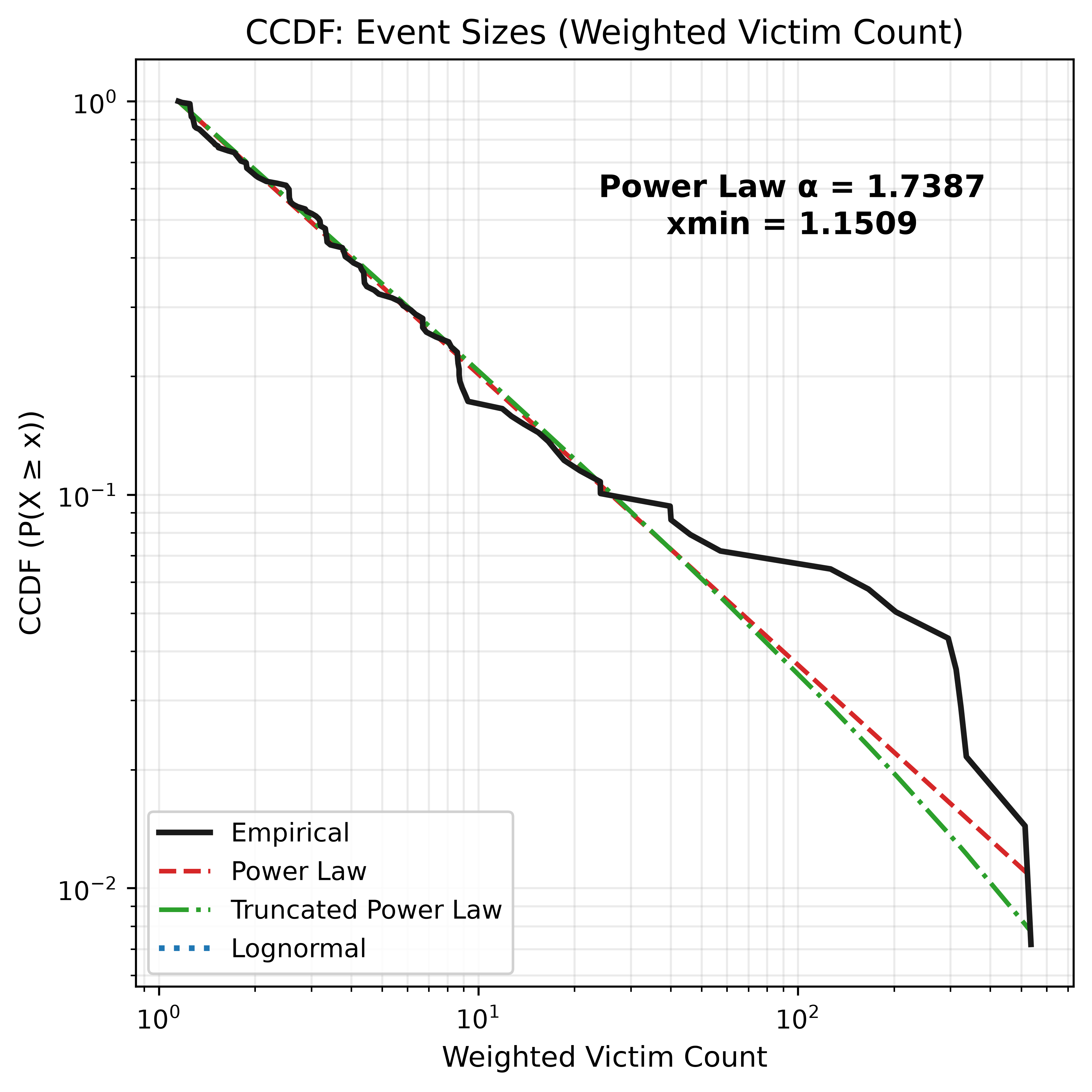}
    \caption{Results using statistical methodology in \cite{Clauset2009} for comparing power law fits to other heavy tailed distributions for cyber kitten weighted victim count.}
    \label{ccdf_single}
\end{figure}

Having laid out our methodology, we present the results in evaluating our cyber kitten distribution for power law plausibility. Our primary result concerning power law plausibility for the distribution of weighted victim count is shown in Figure \ref{ccdf_single}. We measured a p-value score of 0.524, indicating the plausibility of the power law distributions. The power law is best fit with a coefficient value of $\alpha = 1.74$, which is related to the slope $s$ of the CCDF plot simply:

\begin{equation} \label{slope_ccdf} 
s = -(\alpha-1).
\end{equation}

This value can immediately be used to calculate how the probability of larger events scale. To use a simple round number, we can calculate the relative rate of observing an event with 10X magnitude $r_{10X}$ as:

\begin{equation} \label{rate_10x} 
r_{10X} = 10^{s}=10^{-1.74}\approx 0.183
\end{equation}

and a corresponding relative frequency drop $f_{10X}$ as:

\begin{equation} \label{freq_10x} 
f_{10X} = \frac{1}{r_{10X}} \approx 5.5.
\end{equation}

In other words, increasing a cyber kitten's size by 10X results in a frequency that drops by 5.5X.

\subsection{Comparisons to other distributions}\label{distribution_comparisons}
As explained in Section \ref{power_law_testing}, when comparing to power law distributions, negative $R$ values mean the comparison is superior, and a $p$-value $\leq$ 0.05 is considered statistically significant. To apply more rigorous statistics to our results, we performed 10,000 iterations bootstrapping samples with replacement, a standard statistical approach for smaller sample sizes.  The results are shown in Table \ref{tab:dist_comparison}. Power law fits on the bootstrapped samples are consistently plausible, with mean and standard deviation of $p$-value power law fits: 0.4135 ± 0.2363. And in terms of comparison distributions, while both lognormal and power law with cutoff had mean negative $R$ values, the fact that they were both less than -1 indicates a very weak superiority. Similarly, the mean $p$-values are well above the 0.05 threshold. Based on these results, and the goal of this paper to explore worst-case bounds, we continue our final analysis by focusing on pure power laws.

\begin{table*}[htbp]
\centering
\caption{Distribution comparison test results (loglikelihood ratio $R$ and $p$-value)}
\label{tab:dist_comparison}
\begin{tabular}{lcccc}
\toprule
& \multicolumn{2}{c}{$R$} & \multicolumn{2}{c}{$p$-value} \\
\cmidrule(lr){2-3} \cmidrule(lr){4-5}
Compare to & Mean & Std & Mean & Std \\
\midrule
Lognormal            & $-0.638$ & $0.930$ & $0.430$ & $0.394$ \\
Power law with cutoff & $-0.341$ & $0.318$ & $0.483$ & $0.198$ \\
\bottomrule
\end{tabular}
\end{table*}

\subsection{Extrapolation results using best fit power laws}\label{extrapolation}

Having a set of 10,000 best fit power law distributions fit on bootstrapped cyber kitten data, we can finish our analysis by using these best fit functions to extrapolate frequencies for cyber cat magnitudes previously unobserved.

To be precise, for each best fit power law, we use the CCDF function to query probabilities of observing events that are some $M$ multipliers larger than MOVEit, the largest event in our cyber kittens dataset. Defining these query probabilities as $p_q(M)$, where $M \in [1, 10, 100, 250, 1,000]$ (e.g., $M$ = 1 is the same size as MOVEit, $M$ = 1,000 is 1,000 times larger than MOVEit), we can calculate an annual probability $a$ of observing $p_q(M)$ as:

\begin{equation} \label{freq_10x} 
a = n_a p_q(M),
\end{equation}
wherein $n_a$ is the number of cyber events expected per year, given the particular bootstrapped sample and the number of events above the optimal $x_{min}$ value associated with the best fit power law fit. Using $a$, we calculated an expected return period $T$, the number of years expected to observe an event of such magnitude, using the simple equation:

\begin{equation} \label{freq_10x} 
T = \frac{1}{a}.
\end{equation}

The results of this extrapolation analysis are shown in Table \ref{tab:return_period}. As a kind of sanity check, we calculate that on average we would observe a MOVEit-sized event every ~4 years, which is in line with the empirical results that contained 1 such event over a 5 year period of claims. Moving one step further, the results showed that on average, an event 10X MOVEit would be expected every ~28 years. On face value, this seems to be an under-prediction, according to the cat dataset of \cite{johansmeyer2026rewriting}. Using the \$1B loss for MOVEit, there have been 9 such events over the 28 year period from 1998 to 2026. However, as pointed out in the paper itself, 82\% of the aggregate economic losses occurred from 1998--2004, and only 3 events from 2025 to present were 10X MOVEit. Additionally, we have already acknowledged the inexact mapping between victim count and economic loss; while there is a strong correlation, the highest reported economic loss in \cite{johansmeyer2026rewriting} was CrowdStrike at \$1.7B, which ranked 3rd in terms of our weighted victim count. Using this slightly larger economic estimate, only the Worm event from 2005--2026 is 10X larger than CrowdStrike loss. Taking these effects into context, the 10X results seem plausible in relation to the historical cat dataset.

We then explore realms outside of historical precedent, namely losses 100X, 250X, or 1,000X larger than the largest cyber cat events in 2020--2024. The average return period values would indicate that we'd expect such events to happen not on the scale of years or even decades, but on the scale of centuries. Continuing with our focus on analyzing worst-case bounds, we assume that the the economic loss of a MOVEit event can be mapped to \$2B, which is simply rounding up the largest billion from the estimated economic losses from the 4 cyber cats in 2020--2024 in \cite{johansmeyer2026rewriting}. As an example of $M$ = 100, this leads to estimating an upper bound loss of \$200B every 206 years, on average.

\begin{table}[htbp]
\centering
\caption{Power law extrapolation results for expected return period $T$ by loss multiplier $M$.}
\label{tab:return_period}
\begin{tabular}{lcc}
\toprule
$M$ & Mean (years) & Std (years) \\
\midrule
1    & 4.2   & 2.9 \\
10   & 27.8  & 40.2 \\
100  & 205.9 & 621.0 \\
250  & 474.1 & 1904.4 \\
1000 & 906.1 & 4486.4 \\
\bottomrule
\end{tabular}
\end{table}

\section{Discussion}\label{discussion}

As well laid out in \cite{Taleb2007}, true black swan events are, by definition, unpredictable. While cyber insurance could be plagued with an ``unknown-unknown" kind of massive loss event, such a risk will always be (by definition) extremely rare, unpredictable ahead of time, and (arguably) equally likely to happen in virtually any insurance line.

This work supports the plausibility that systemic cyber risk follows a power-law like pattern observed in many different perils, moving cyber risk deeper into the ``gray swan" territory, a world wherein one possesses the ability to make extrapolations on previously unseen events based on an underlying law that dictates the system. Have we seen the largest cyber event we will ever see? Almost certainly not. Can we predict exactly when such an event will happen? Again; no. But the power of having a gray swan vs a black means there is meaningful information concerning the potential magnitude and frequency of events which can take place.

Based on the data in this research, we see several encouraging observations that would support cyber being a more acceptable risk as a line of insurance than typical perceptions and media would convey. The worst-case bounding exercise would calculate that a loss of \$200B or higher would be expected once every 200+ years. A loss of this magnitude is massive but by no means crippling to the global insurance market, nor unprecedented. As noted in \cite{johansmeyer2026gdp}, natural perils dwarf the kinds of losses so far seen in the cyber space in terms of pure economic loss. Over just the last 20 years, there have been 5 natural disasters costing over \$100B (adjusted for inflation), 2 of which were over \$200B. These events were born by the insurance market, and findings from this research would suggest that comparably sized cyber losses would be expected at far greater return periods.

This work has also explicitly focused on the worst-case scenario situation for cyber, but there is also reason to believe this is overly pessimistic. An optimistic interpretation of cyber risk would assert that feedback loops limit worst case damage potential. Feedback loops are processes where the product of a system becomes part of the same system's input, leading to self-regulation and/or reinforcement. Coming back to our original example in Section \ref{introduction}, an example of such loops in relation to a real-world risk was observed during the COVID-19 pandemic. Many groups attempted to model the spread, with some, such as the UK's Warwick University model, significantly over-predicting actual infections \cite{Sayers2022}. A wide ranging study of 61 different Covid models in \cite{Ioannidis2022} showed that the simplest explanation between ``good" and ``bad" models was that the bad ones failed to incorporate adaptive behavior into the equation (i.e., feedback loops). Very naively, higher death rates naturally leads to more news coverage and higher perceived risk, which then translates into lower transmission rates. This kind of simple feedback loop added to a simple SEIR model (susceptible, exposed, infected, recovered) beat virtually all other models in study without one, and could predict considerably better for longer time horizons. 

There are various examples of how these feedback mechanisms play out in cyber risk. \cite{Algarni2021} supports the data breaches have a kind of economy of scale, and loss per record declines the larger the size of the breach. Another case study of feedback loops in action was detailed in \cite{Greenberg2018}, concerning the NotPetya attack. As word spread from the growing number of victims, unaffected companies started powering down computers manually to avoid further infections. An interesting future research direction would be to experiment with different kinds of feedback functional forms that still have underlying power law-like structure.

Finally, the work here supports a modeling assumption that cyber cat events follow a power law-like distribution, evidenced by cyber claims data. This assumption may rightly evoke criticism from the broader cyber community, who in general approach cyber risk much differently. For an individual company, cyber risk is specific: it matters which technologies are impacted by the newest zero-day vulnerability, whether your businesses would be hurt more by interrupting real time production or by the liability of a data breach, and what specific coverage areas are in your cyber insurance policy. But as highlighted in Section \ref{methodology}, one of the beauties of power laws is that their intrinsic nature rejects the near infinite details, data, and context that could inform a more complex model. Rather, it asserts a simplified relationship between a small number of macroscopic properties, that accurately predict a macroscopic result. We believe that the capability of modeling the relationship between worst-cast cyber cat magnitude and frequency relationships by itself is a useful direction for the field, and encourage other carriers and re-insurers to run similar analysis to validate the observed power law behavior, as broader validation of cyber's apparent stability as a risk class. 

\section{Acknowledgments}\label{acknowledgments}
We'd like to thank the many reviewers who helped guide the final version of this work, including David Ross and Simon Welton. In particular, we'd also like to thank the holistic guidance of Tom Johansmeyer, which substantially refined the key story told in this work. LLM technology was not used in writing any of the prose of this paper, but was used for formatting / gathering citations, reducing abstract word count, and the explicit cyber kittens enrichment discussed in length in this piece.

\printbibliography

\end{document}